\documentclass[oneside,twocolumn,epsfig]{revtex4}
\usepackage{graphicx}

\begin{document}
\title[] {Dynamical Stochastic Processes of Returns in Financial Markets
 }
\author{Gyuchang \surname{Lim$^{1}$}}
\author{SooYong \surname{Kim$^{1}$}}
\author{Junyuan \surname{Zhou$^{1}$}}
\author{Seong-Min \surname{Yoon$^{2}$}}
\author{Kyungsik \surname{Kim$^{3}$}}
\affiliation{$^{1}$Department of
Physics, Korea Advanced Institute\\
of Science and Technology, Daejeon 305-701, Korea\\
$^{2}$Division of Economics, Pukyong National University,\\
Pusan 608-737, Korea\\
$^{3}$Department of Physics, Pukyong National University,\\
Pusan 608-737, Korea\\
}

\begin{abstract}

We study the evolution of probability distribution functions of
returns, from the tick data of the Korean treasury bond (KTB)
futures and the S$\&$P $500$ stock index, which can be described by
means of the Fokker-Planck equation. We show that the Fokker-Planck
equation and the Langevin equation from the estimated Kramers-Moyal
coefficients are estimated directly from the empirical data. By
analyzing the statistics of the returns, we present quantitatively
the deterministic and random influences on financial time series for
both markets, for which we can give a simple physical
interpretation. We particularly focus on the diffusion coefficient
that may be significantly important for the creation of a
portfolio.\\
\hfill\\
PACS number(s): 05.45.-a, 05.10.-a, 05.40.-a, 89.75.Hc\\
\end{abstract}

\maketitle

\section { Introduction }

In the pioneering work of Bachelier $[1]$, it has been shown that
the probability of price changes can be determined in terms of the
Chapman-Kolmogorov equation and the Wiener process. Since this study
was published, many scientific researchers in this area  have
suggested the stochastic models to explain how the probability
density function (PDF) in delay times behaves dynamically on
rigorous statistical principles. To find the scale invariance and
universality in the stochastic model is an important and powerful
issue in the statistical analysis of tick data of various financial
markets $[2,3]$. Each financial market has different characteristic
features but they are expected to have some similar and
scale-invariant properties in common. We wish to focus on the
findings of unique trends and on the universality in them. In this
study, we mainly analyze two different financial markets, the KTB
and the S$\&$P $500$ stock index. The former is supposed to open for
a short term of six months while the latter continues without breaks
except for holidays. It is furthermore shown that there exists a
similarity in the correlation of returns for both of these markets.

Several studies on the dynamics of financial markets have recently
focused on the understanding of the fat-tailed PDF of the price
changes. It has been shown from previous works $[4,5]$ that price
changes in the empirical dollar-mark exchange rates is regarded as a
stochastic Markov process for different delay times. Ghashghaie et
al. $[6]$ have discussed an analogy between the dynamical
high-frequency behavior of the foreign exchange market and the
short-time dynamics of hydrodynamic turbulence, and they are shown
to be similar to the energy cascade from large and small time scales
in the theoretical model of turbulence $[7,8]$. Furthermore, Ivanova
et al. $[9]$ showed the validity of the Fokker-Planck approach by
testing the various financial markets, i.e., NASDAQ, NIKKEI 225, and
JPY/DEM and DEM/USD currency exchange rates.

Motivated by the work of Friedrich and Ivanova $[4,5,9]$, we apply
the Fokker-Planck equation approach to two sets of financial data,
in this case the KTB and the S$\&$P $500$ stock index. In this
paper, we derive the functional form of the Fokker-Planck equation
from the estimated Kramers-Moyal coefficients and present an
equivalent Langevin equation for each. In Section $2$, we present
the numerical method for estimating Kramers-Moyal coefficients from
price returns of the real financial time series. We present some
results obtained by the numerical simulations in Section $3$, and
our concluding remarks are in the final section.

\section { The Fokker-Planck equation and the Langevin equation}

In our study the return $r(t)$, given by $r(t)=\ln[p(t+\Delta
t)/p(t)]$ for the price $p(t)$ at time $t$, is assumed to follow a
Markov process. This assumption becomes plausible by the validity of
the Chapman-Kolmogorov equation. To make a statistical analysis of
financial markets, we consider a $N$-point joint PDF, $p(r_N ,
\Delta t_N ;r_{N-1} , \Delta t_{N-1};...; r_1 , \Delta t_1 )$.
However, in the case of a Markov process, it is sufficient to
consider the joint PDF generated by a product of the conditional
probabilities in the following way:
\begin{eqnarray}
&&p(r_N , \Delta t_N ;...; r_1 , \Delta t_1) \nonumber\\
&&=p( r_N , \Delta t_N | r_{N-1} , \Delta t_{N-1} )\cdot\cdot\cdot
p( r_2 , \Delta t_2 | r_1 , \Delta t_1 ), \nonumber\\
\label{eq:a1}
\end{eqnarray}
%
%
where the conditional probability $p( x_{2} , \Delta t_{2} | x_1,
\Delta t_1 )$ is defined by
\begin{equation}
p( r_{2} , \Delta t_{2} | r_1, \Delta t_1 )= p( r_{2} , \Delta t_{2}
; r_1, \Delta t_1 ) /p( r_{1} , \Delta t_{1}) \label{eq:a22}
\end{equation}
and evaluated for the time delay $\Delta t_2 < \Delta t_1$ directly
from the tick data set. In addition we consider the
Chapman-Kolmogorov equation as a necessary condition
\begin{equation}
p( r_2 , \Delta t_2 | r_1 , \Delta t_1 )= \int dr_i p( r_2 , \Delta
t_2 | r_i , \Delta t_i )p( r_i , \Delta t_i | r_1 , \Delta t_1 ),
\label{eq:a3}
\end{equation}
which holds for any time lag $\Delta t_i $, with $\Delta t_2 <
\Delta t_i < \Delta t_1$.

\begin{figure}[]
\includegraphics[angle=0,width=9.5cm]{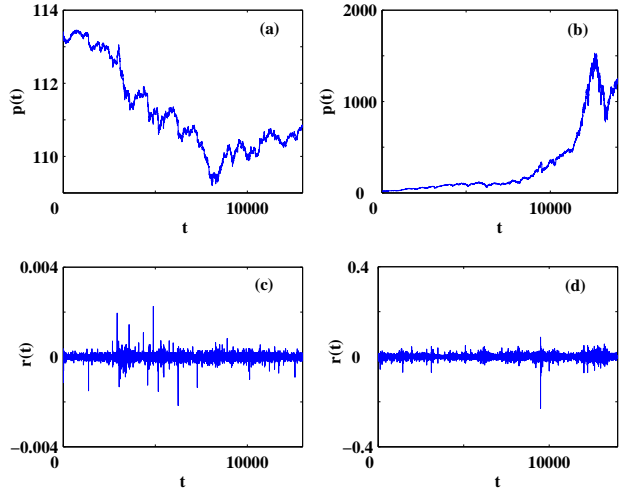}
\caption[0]{Plot of the time series and returns in the KTB$503$
((a)and (c)) and the S$\&$P $500$ stock index ((b)and (d)).}
\end{figure}
\begin{figure}[]
\includegraphics[angle=0,width=8.5cm]{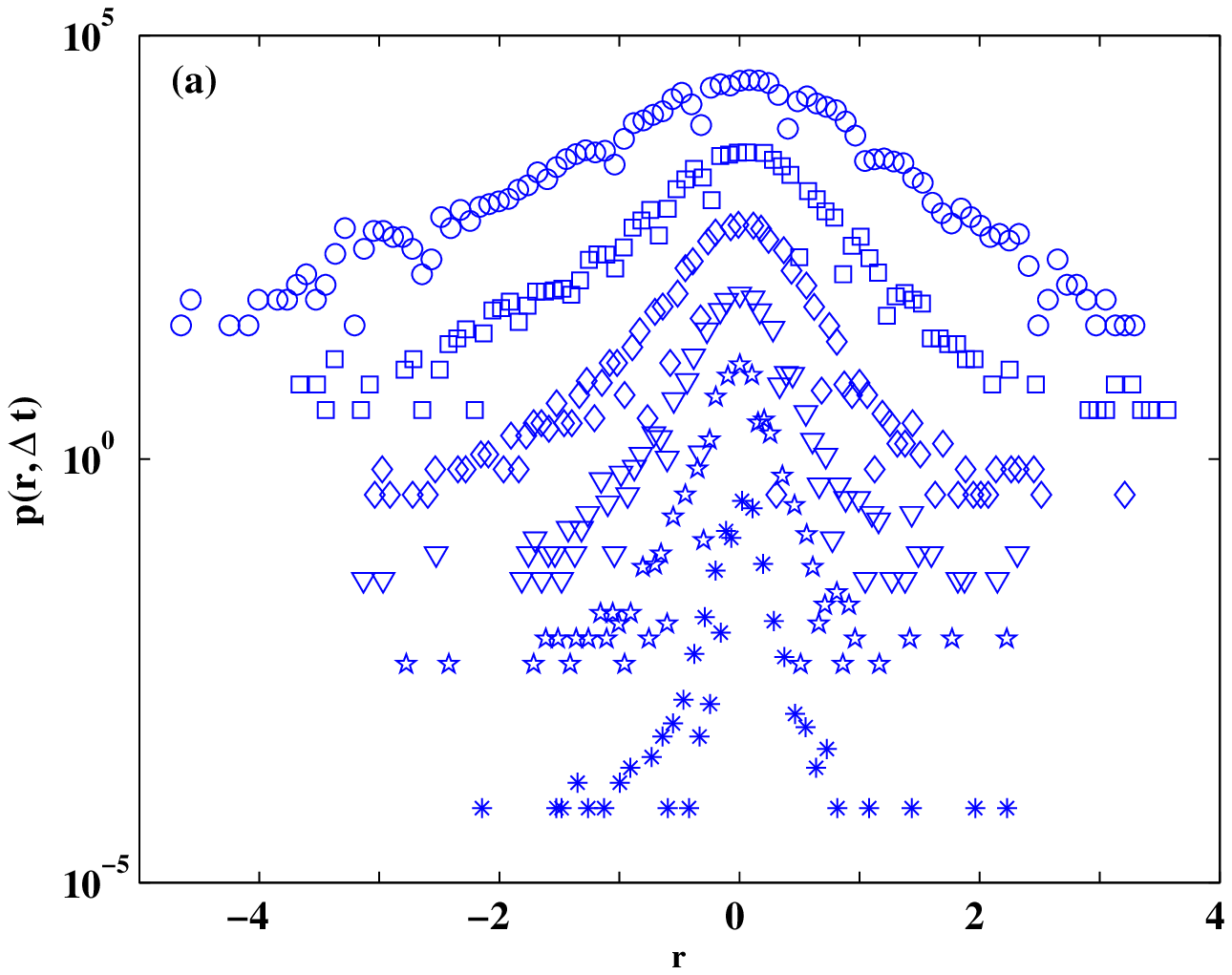}
\includegraphics[angle=0,width=8.5cm]{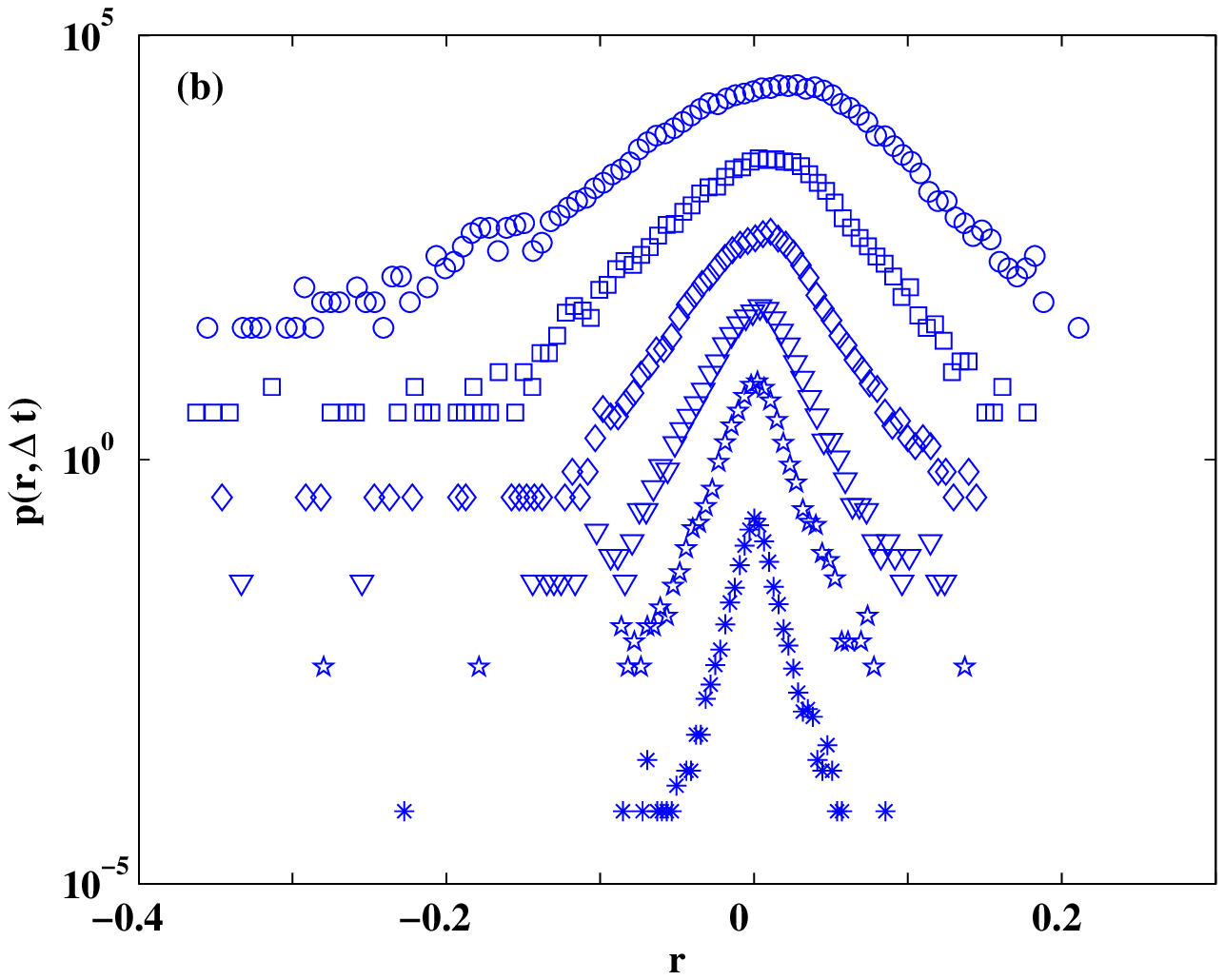}
\caption[0]{Plot of the PDFs of returns for time lags $\Delta
t=1,2,4,16, 32$ min and day (from bottom to top) in (a) the KTB$503$
and (b) the S$\&$P $500$ stock index. The top figures (circles) of
(a) and (b) approximately approach to the Gaussian PDF.}
\end{figure}

In order to derive an evolution for the change of distribution
functions across the time scales, we should show the validity of the
Chapman-Kolmogorov equation.
From the validity of the
Chapman-Kolmogorov equation, we can obtain an effective
Fokker-Planck equation. This equation is represented in terms of the
differential form as follows:
\begin{equation}
\frac{d}{d\Delta t} p( r,\Delta t) = [- \frac{\partial}{\partial r}
D_{1} ( r,\Delta t)+\frac{{\partial}^{2}}{{\partial} r^2} D_{2} (
r,\Delta t)] p( r,\Delta t), \label{eq:a4}
\end{equation}
where the drift coefficient $D_{1} ( r,\Delta t)$  and diffusion
coefficient $D_{2} ( r,\Delta t)$ can be estimated directly from
moments $M_{k}$ of the conditional probability. The relations
between coefficients and moments in Kramers-Moyal expansion are
described as follows:
\begin{equation}
D_{k}(r,\Delta t)=\frac{1}{k!} \lim_{{\Delta\tau} \rightarrow
0}M_{k}, \label{eq:a5}
\end{equation}
\begin{equation}
M_{k}=\frac{1}{\Delta t} \int {dr^{'}}(r^{'} - r)^{k} p( r^{'}
,t+\Delta t | r,t) \label{eq:a6}
\end{equation}
Therefore we can derive a Lagevin equation from Eq. $(4)$.
\begin{equation}
\frac{d}{d\tau} r(\tau)=D_{1} ( r(\tau),\tau)+\sqrt{D_{2} (
r(\tau),\tau)}f(\tau), \label{eq:a9}
\end{equation}
where $f(\tau)$ is a fluctuating $\delta$-correlated white noise
with a Gaussian distribution holding for $\langle
f(\tau)f({\tau}^{'})\rangle=2\delta(\tau-{\tau}^{'})$.

\section {Numerical calculations}

\begin{figure}[]
\includegraphics[angle=0,width=8.5cm]{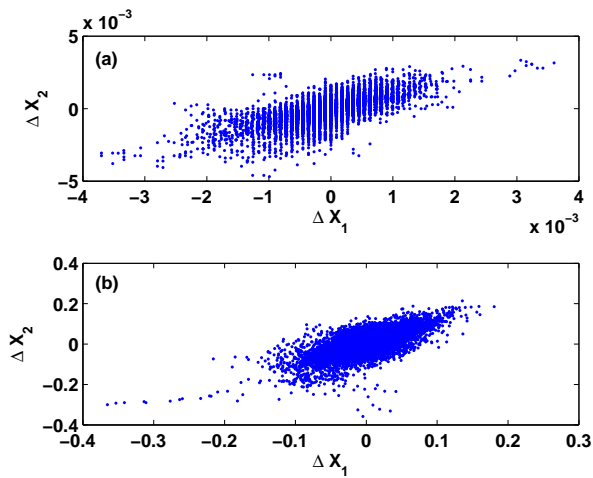}
\caption[0]{Contour plot of the joint PDF $p( r_{2} , \Delta t_{2} ;
r_{1}, \Delta t_{1} )$ for the simultaneous occurrence of price
returns with time scales $\Delta t_{2}=32$ (min, day) and $\Delta
t_{1}=16$ (min, day) in (a) the KTB$503$ and (b) the S$\&$P $500$
stock index, respectively.}
\end{figure}
\begin{figure}[t]
\includegraphics[angle=0,width=8.5cm]{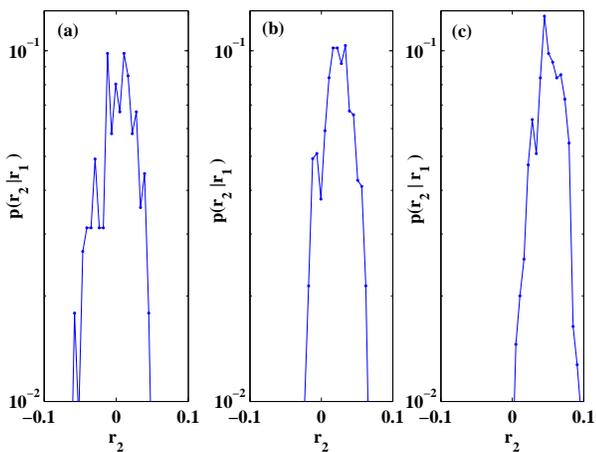}
\caption[0]{Contour plot of the conditional PDF $p(r_{2} , \Delta
t_{2} |r_{1} , \Delta t_{1} )$ for S$\&$P $500$ stock index with $
\Delta t_{1} =16$ and $\Delta t_{2} =32$ days. (a), (b), and (c) are
corresponding cuts for $r_1$ = $- 0.05$, $0$, and $0.05$}
\end{figure}
\begin{figure}[t]
\includegraphics[angle=0,width=9.0cm]{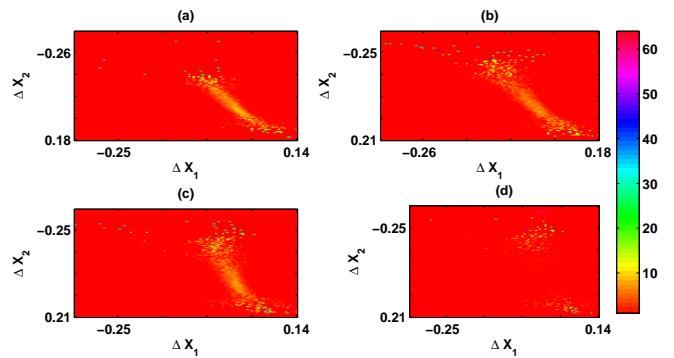}
\caption[0]{Matrix representations of the conditional PDFs of S$\&$P
$500$ stock index scaled in a jet colorbar. (a), (b), and (c)
correspond to $p(r_{2} , \Delta t_{2} |r_{1} , \Delta t_{1} )$ with
each $(\Delta t_{1} ,\Delta t_{2} )$ set to $(8,16), (16,32),$ and
$(8,32)$ in days in sequence. (d) indicates the different
conditional PDF between the r.h.s and the l.h.s. of the
Chapman-Kolmogorov equation, Eq. $(3)$.}
\end{figure}
\begin{figure}[t]
\includegraphics[angle=0,width=8.5cm]{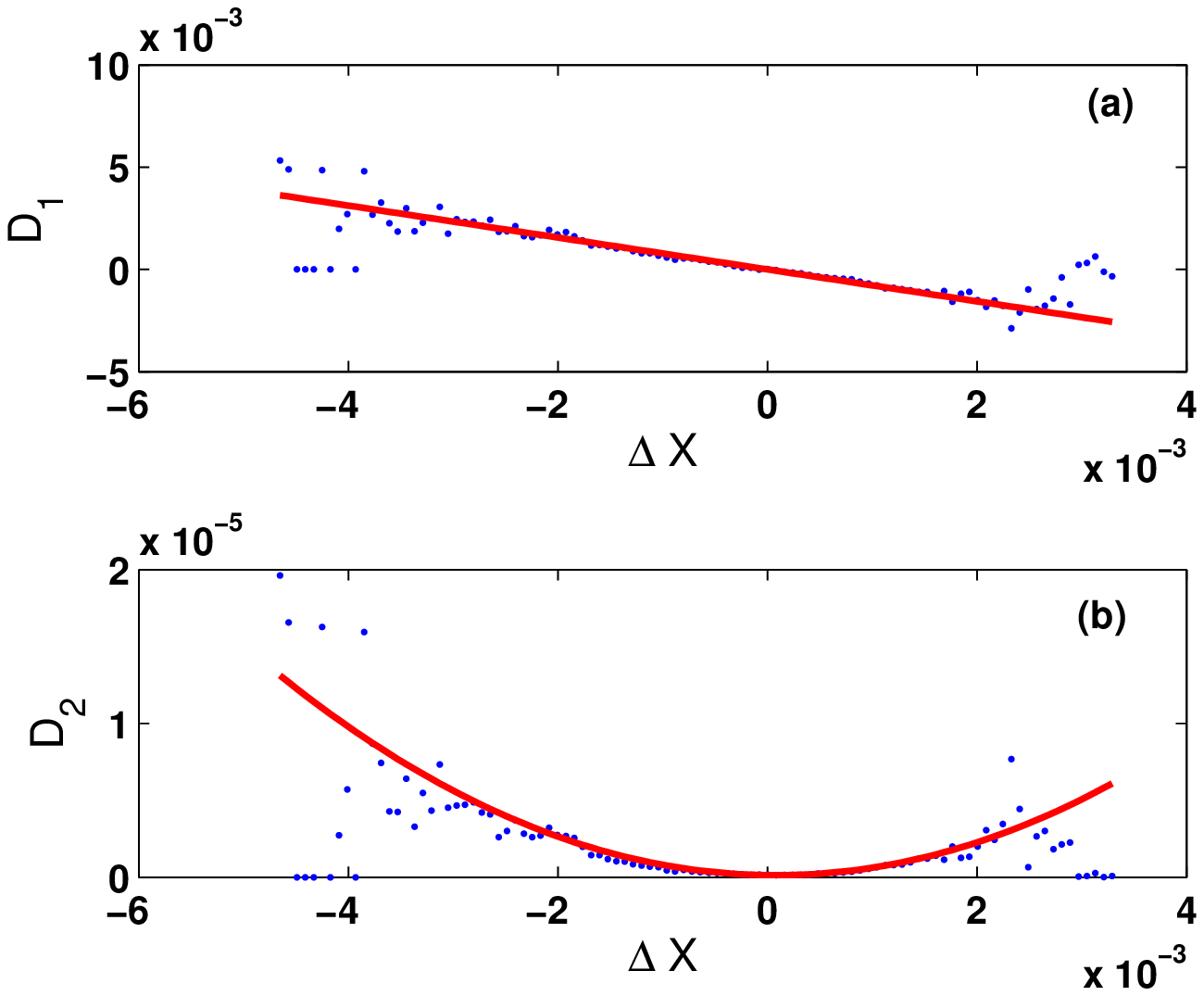}
\includegraphics[angle=0,width=7.5cm]{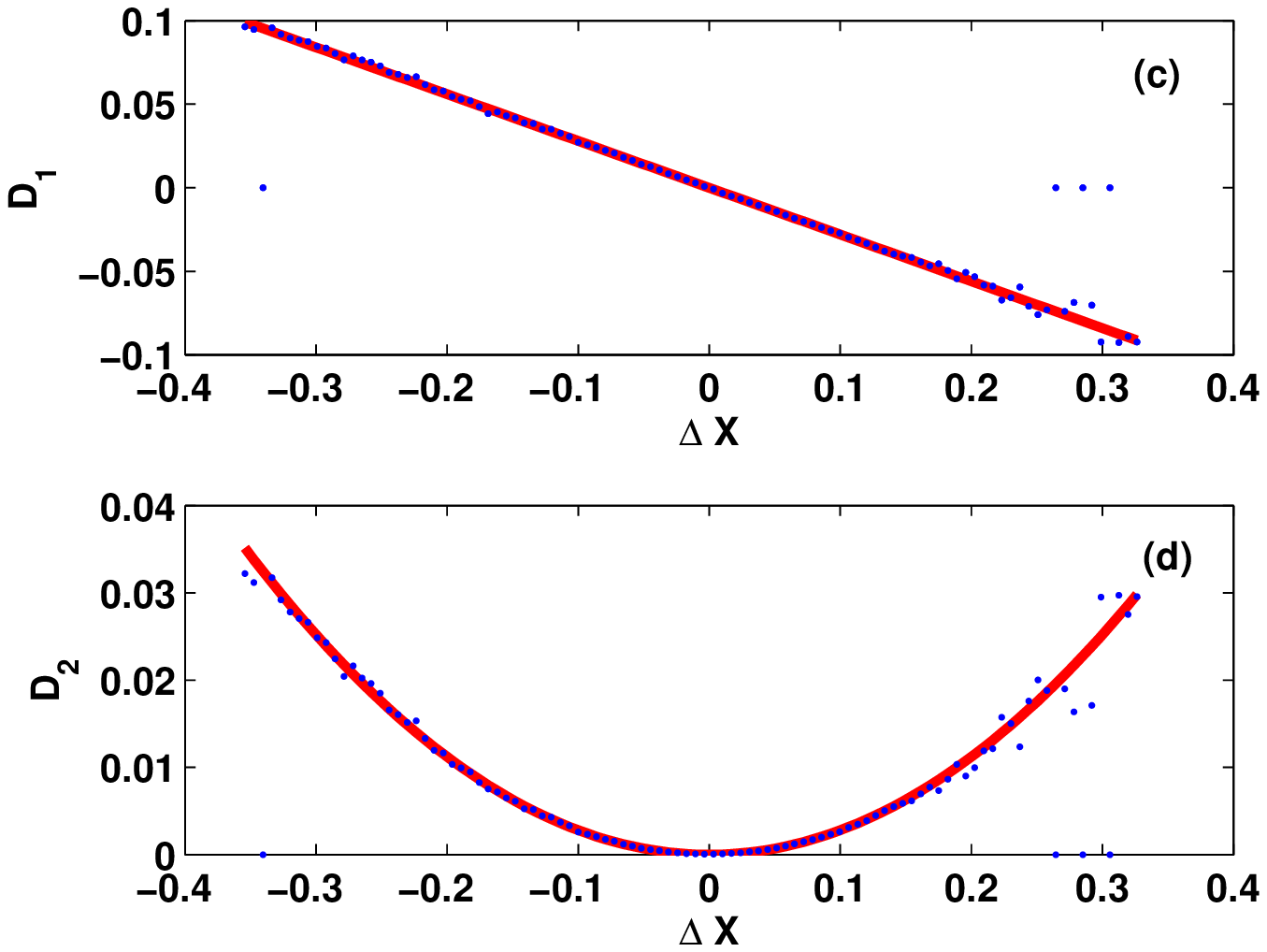}
\caption[0]{Drift and diffusion coefficients estimated from the
conditional PDFs. The solid curves present linear and quadratic fits
in the KTB$503$ ((a) and (b)) and S$\&$P $500$ stock index ((c) and
(d)).}
\end{figure}
\begin{figure}[t]
\includegraphics[angle=0,width=8.5cm]{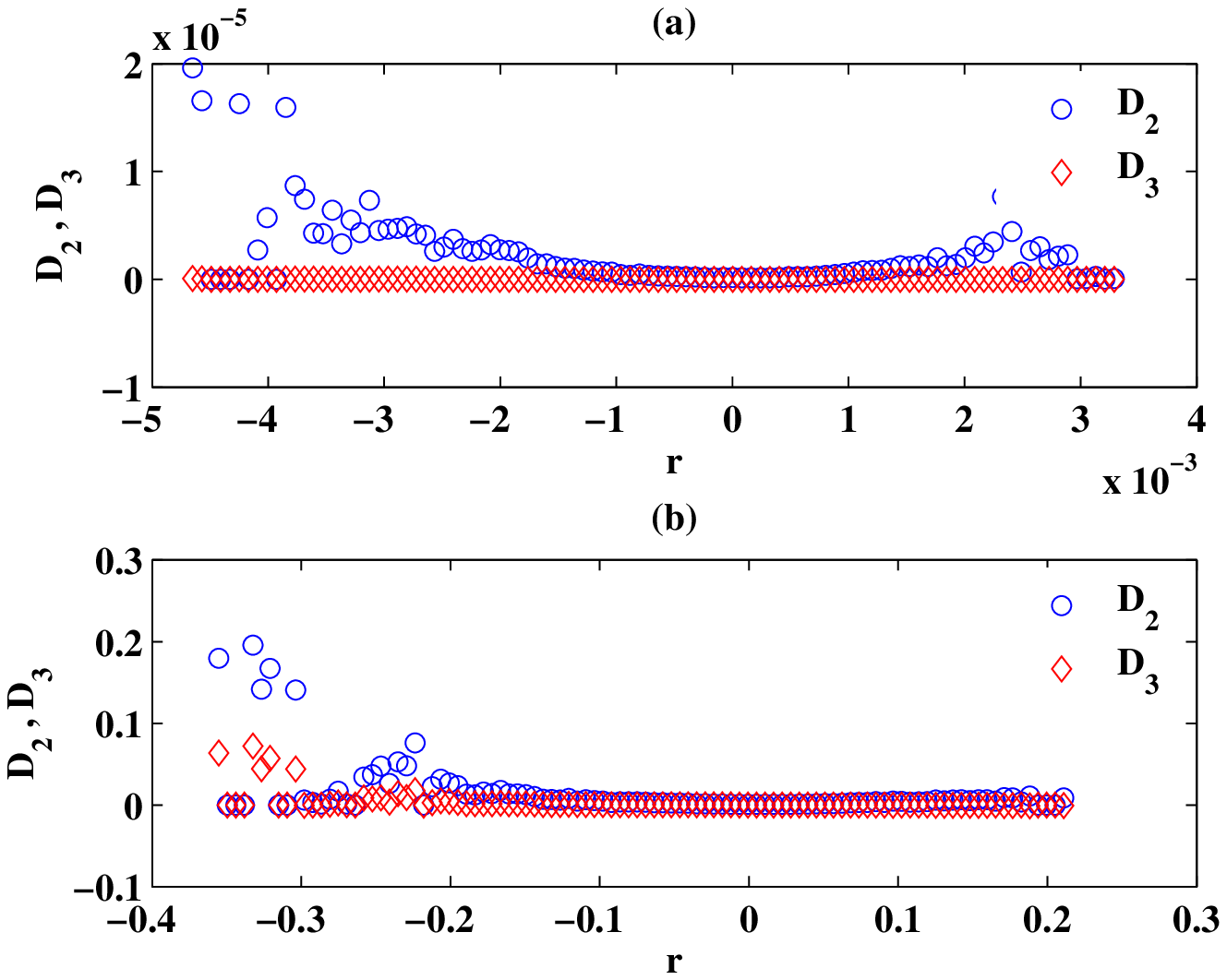}
\caption[0]{Plot of $D_3$(open circle) vs $D_2$(diamond). (a) and
(b) correspond to the KTB$503$ and S$\&$P $500$ stock index,
respectively.}
\end{figure}

We analyze two different databases, the KTB $503$ and the S$\&$P
$500$ stock index. The KTB $503$ is a minutely tick-by-tick data set
consisting of $13282$ quotes taken from $1$st October $2004$ to
$31$st March $2005$, and the S$\&$P $500$ stock index is a daily
data set consisting of $14027$ quotes taken from $1950$ to $2005$.
From both data sets we measure the price returns $r(t)$ for
analysis, which is very useful to see the gain or loss of an asset
in a given time period.
%
%
Both time series are plotted together with the returns of the time
delay $\Delta t_{max}=32$ minutes and $32$ days for the  KTB $503$
and S$\&$P $500$ stock index, respectively in Fig. $1$. We consider
a logarithmic time scale $\tau = \ln (\Delta t_{max} /\Delta t)$ for
the sake of convenience in the analysis.

We show the empirical PDF for various time delays in Fig. $2$ from
which we can confirm the tendency for the PDF to approach a Gaussian
PDF as the time lag $\Delta t$ increases. In a comparison with the
PDF of Ref. $[4]$, it is shown that the numerical iteration of the
effective Fokker-Planck equation fits well the empirical PDF. In
this paper, we restrict ourself to obtain an effective Fokker-Planck
equation and to confirm that the approach presented is also well
applied to the other financial markets different from the foreign
exchange market. For the following analysis, we calculate
conditional probabilities by using the contour plot of the joint PDF
for the simultaneous occurrence of price returns, as shown in Fig.
$3$.

We can obtain approximately conditional probabilities by counting
the simultaneous events belonging to each bin by which the
corresponding events can be divided (see Fig. $4$).

Fig. $5$ shows the matrix representation of conditional
probabilities, which is a good indicator for proving the validity of
the Chapman-Kolmogorov equation for the empirical data. The matrix
is generated from a histogram with $100$ discretionary steps over
the range of $r(t)$. There are visible deviations in the outer
region of returns, which are probably due to the finite resolution
of the statistics and to the small number of data set, where $\Delta
t_{max}=32$ days and $32$ minutes for S$\&$P$500$ stock index and
the KTB$503$, respectively.

To calculate $M_{k}$ approximately we apply the conditional PDF
obtained from the procedure, as shown in Figs. $3$ and $4$. We set
$\Delta \tau = \ln{2}$ in the estimation of $D_{k}$, and the
functional $r$ dependence of $D_{1}$ and $D_{2}$ is shown in Fig.
$6$. We confirmed that there is no significant difference in the
estimated coefficients when varying $\Delta t_{max}$, provided that
the $\Delta t$ is fixed. For all scales $ t$ and $\Delta t $, the
functional dependence of $ D_{1}$ and $D_{2}$ behaves in the same
manner. $D_{1}$ shows a linear dependence on returns, while $D_{2}$
is approximated by a polynomial of two degrees in the logarithmic
returns. As specified in Ref. $[4]$ we add an exponential term to
correct the delicate differences in the minimum values of $D_{2}$
although this is too small to be considered significant. By
analyzing the plots in Fig. $6$, we obtained the following
approximations for the KTB$503$ and S$\&$P $500$ stock index:
For S$\&$P $500$ stock index
\begin{equation}
D_{1}=-0.70 r \label{eq:a7}
\end{equation}
and
\begin{equation}
D_{2}=0.48( r-5.06\times10^{-3})^2+7.63\times10^{-4}\exp(-1.01\tau).
\label{eq:a71}
\end{equation}
For the KTB$503$
\begin{equation}
D_{1} = -0.78 r \label{eq:a72}
\end{equation}
and
\begin{equation}
D_{2}=0.58(r-8.21\times10^{-5})^{2} +1.31\times10^{-7}
\exp(-4.46\tau). \label{eq:a8}
\end{equation}

In Eqs. $(9)$ and $(11)$, we obtained the exponential term by
solving the simultaneous equations from a pair of two conditional
conditions, in this case $p(r_{2}, \Delta t_{2} | r_{1} ,\Delta
t_{1} )$ where $\Delta t_{1}$ and $\Delta t_{2}$ are chosen to
satisfy the condition, $\tau=\ln{2}$. The analogy between turbulence
and the foreign exchange market was well described and argued in
Ref. $[4]$. Here we extend the argument to the stock and futures
exchange markets. Although, as mentioned previously, the KTB$503$
has a short-term expiration different from the S$\&$P$500$ stock
index, they both show good agreement in the form of the
Kramers-Moyal coefficients. We here stress again that the
Fokker-Planck approach can be a suitable method to analyze the
financial tick data and in understanding the statistics of their
returns. Furthermore, we present a nonlinear Langevin equation, here
an underlying stochastic process, by showing that the Kramers-Moyal
coefficients vanish for $k\geq3$ (see Fig. $7$). It can be
particularly confirmed that the third moment of Kramers-Moyal
coefficients vanishes. In Eq. $(9)$ and $(11)$, the exponential
terms are too small to be negligible when compared to other terms.
Hence, from the estimated results we can obtain a linear stochastic
equation with multiplicative noise, as a quadratic noise
Ornstein-Uhlenbeck process, as follows:
For S$\&$P $500$ stock index,
\begin{equation}
\frac{d}{d\tau} r(\tau) = -0.70  r(\tau)+\sqrt{0.48} r(\tau)f(\tau).
\label{eq:a10}
\end{equation}
For the KTB$503$,
\begin{equation}
\frac{d}{d\tau} r(\tau)=-0.78 r(\tau)+\sqrt{0.58} r(\tau)f(\tau).
\label{eq:b15}
\end{equation}
From Eqs. $(8)$-$(11)$ we can see that the KTB$503$ has a greater
drift and diffusion coefficients than the S$\&$P$500$ stock index.
This implies that the former is more dynamic than the latter.
Furthermore, the volatility of returns has considerable influence
upon the deterministic trend in that the diffusion coefficient $D_2$
is somewhat comparable to the drift coefficient $D_1$.


\section {Conclusions}

We have showed that the Fokker-Planck approach can be a good method
for the analysis of the financial markets in a wide spectrum. Even
for a short-term expired financial market such as the Korean
treasury bond futures we could obtain a well-fitted result
describing the behavior of its PDF in time delays. We have also
obtained the same stochastic process from both analyzed financial
markets irrespective of their data acquisition frequency.
Quantitative measurements of deterministic and random influences for
both futures and stock markets imply that the futures market is more
dynamic than the stock market.



This paper has shown that the temporal correlation of the futures
market can be described well in terms of Fokker-Planck equations.
Particularly, the KTB$503$ is transacted for a short term of six
months in the Korean financial market, but it is striking that its
functional form of the Kramers-Moyal coefficients is approximately
consistent with that of other assets. Moreover, the comparison of
the drift and diffusion coefficients with various financial markets
provides us with a numerical indicator for the creation of
portfolios and risk management strategies. In the future, it will be
of interest to compare our results with those of other national
options. Thereby we can measure and compare the stability and the
efficiency of various financial markets.

%
\begin{acknowledgements}
This work was supported by the Korea Research Foundation Grant
funded by the Korean Government (MOEHRD) (KRF-2005-041-C00183).

\end{acknowledgements}


\begin{thebibliography}{}

\bibitem{Ba1} L. J. B. Bachelier, \emph{Theorie de la Speculation}, (Gauthier-Villars, Paris,
1900); reprint (Editions Jaques Gabay, Paries, 1995).
\bibitem{St2} H. E. Stanley, \emph{Proceedings of the 22nd IUPAP International Conference on Statistical Physics},
S. Dattagupta, H. R. Krishnamurthy, R. Pandit, T. V. Ramakrishnan
and D. Sen (ed.), Indian Academy of Sciences, Bangalore, 2004, pp.
645-660.
\bibitem{Ma3} R. N. Mantegna and H. E. Stanley, $An$ $Introduction$ $to$ $Econophysics$:
$Correlation$ $and$ $Complexity$ $in$ $Finance$ (Combridge
University Press, Cambridge, 2000).
\bibitem{Fr4} R. Friedrich, J. Peinke and Ch. Renner, Phys. Rev. Letter {\bf 84}, 5224
(2000).
\bibitem{Fr5} Ch. Renner, J. Peinke and R. Friedrich, Physca {\bf A298},
499 (2001).
\bibitem{Ga6} S. Ghashghaie, W. Breymann, J. Peinke, P. Talkner and
Y. Dodge, Nature {\bf 381}, 767 (1996).
\bibitem{Da7} J. Davoudi and M. R. R. Tabar, Phys. Rev. Letter {\bf 82}, 1680
(1999).
\bibitem{Fr8} U. Frisch, \emph{Turbulence} (Cambridge University
Press, Cambridge, England, 1995).
\bibitem{Iv9} K. Ivanova, M. Ausloos and H. Takayasu, \emph{The Application of Econophysics},
H. Takayasu (Ed.), Springer, Tokyo, 2004, pp. 161-168.


\end{thebibliography}
\end{document}